\documentclass[11pt]{article}

\usepackage[a4paper,top=2.5cm,bottom=2.cm,left=2.8cm,right=2.5cm,marginparwidth=1.75cm]{geometry}

\usepackage{authblk}
\usepackage{lineno}
\usepackage{amssymb}
\usepackage{amsmath}
\usepackage{amsthm}
\usepackage{epsfig}
\usepackage{graphicx}
\usepackage{graphics}
\usepackage{float}
\usepackage{multirow}
\usepackage{color}
\usepackage{lineno}
\usepackage[normalem]{ulem} 
\usepackage{makeidx}
\usepackage{xspace}
\usepackage{cite}

\usepackage[colorlinks=true, linktoc=page, linkcolor=blue, citecolor=blue, urlcolor=blue]{hyperref}

\newcommand{\pp}{\mathrm{pp}}
\newcommand{\ppbar}{\mathrm{p\bar{p}}}
\newcommand{\PbPb}{\mathrm{PbPb}}

\newcommand{\epem}{e^+e^-}
\newcommand{\gaga}{\gamma\gamma}

\newcommand{\pT}{p_{\rm T}}
\newcommand{\ET}{E_{\rm T}}

\newcommand{\sqrts}{\sqrt\mathrm{s}}
\newcommand{\sqrtsnn}{\ensuremath{\sqrt{s_{_{NN}}}}}

\newcommand{\starlight}{\textsc{Starlight}}

\newcommand*{\cm}{c.m.\@\xspace}
\newcommand*{\eg}{e.g.\@\xspace}
\newcommand*{\ie}{i.e.\@\xspace}

\usepackage{caption} 
\begin{document}

\vspace{-0.2cm}

\title{Collider constraints on axion-like particles}

\author[1]{David~d'Enterria\footnote{On behalf of the ATLAS and CMS Collaborations. E-mail: \href{mailto:dde@cern.ch}{david.d'enterria@cern.ch}}
\vspace{-0.2cm}}
\affil[1]{CERN, EP Department, CH-1211 Geneva, Switzerland}

\setcounter{page}{1}

\date{}
\maketitle

\vspace{-1.0cm}
\begin{abstract}
The current status and future prospects of searches for axion-like particles (ALPs) at colliders, mostly focused on the CERN LHC, are summarized. Constraints on ALPs with masses above a few GeV, that couple to photons, as well as to Z or Higgs bosons, have been set at the LHC through searches for new $a\to\gaga$ resonances in di-, tri-, and four-photon final states. Inclusive and exclusive diphotons in proton-proton and lead-lead collisions, $\pp,\,\PbPb \to a \to \gaga\,(+X)$, as well as exotic Z and Higgs boson decays, $\pp \to \mathrm{Z},\mathrm{H}\to a\gamma \to 3\gamma$ and $\pp \to \mathrm{H}\to aa \to 4\gamma$, have been analyzed. Exclusive searches in PbPb collisions provide the best exclusion~limits for ALP masses $m_a\approx 5$--100~GeV, whereas the other channels are the most competitive ones over $m_a\approx 100$~GeV--2.6~TeV. Integrated ALP production cross sections up to $\sim$100~nb are excluded at 95\% confidence level, corresponding to constraints on axion-photon couplings down to $g_{a\gamma}\approx$~0.05~TeV$^{-1}$, over broad mass ranges.
Factors of 10--100 improvements in these limits are expected at the LHC approaching $g_{a\gamma}\approx 10^{-3}$~TeV$^{-1}$ over $m_a\approx 1$~GeV--5~TeV in the next decade. 
\end{abstract}

\section{Introduction}

The existence of fundamental scalar particles in nature has received very strong evidence with the discovery of the Higgs boson. Searches for additional (pseudo)scalar particles have thereby been attracting an increasing interest in the last years in collider studies of physics beyond the Standard Model (BSM)~\cite{Jaeckel:2012yz,Mimasu:2014nea,Jaeckel:2015jla,Bauer:2017ris}, and have become key ingredients of the ``intensity frontier'' physics program aiming at discovering new feebly interacting particles~\cite{Essig:2013lka,Lanfranchi:2020crw}. Axion-like particles (ALPs) are hypothetical pseudoscalar bosons that appear naturally in many extensions of the SM, often as pseudo Nambu--Goldstone bosons (pNGBs) arising in the spontaneous breaking of a new global approximate symmetry~\cite{Ringwald:2014vqa} 
(similarly as the neutral pion does for the chiral symmetry of quantum chromodynamics, QCD).
ALPs are ubiquitous in a variety of particle physics and cosmology BSM scenarios such as
\begin{description}
\item (i) the original axion, a pNGB of the Peccei--Quinn symmetry broken by the axial anomaly of QCD, introduced to solve the absence of observed charge-parity (CP) violation in the strong interaction (``strong CP problem'')~\cite{Peccei:1977hh,Weinberg:1977ma,Wilczek:1977pj};
\item (ii) light pseudoscalars proposed as cold dark matter (DM) candidates particles~\cite{Duffy:2009ig,Marsh:2015xka}, or dark-sector mediators~\cite{Nomura:2008ru,Dolan:2014ska,Kozaczuk:2015bea}; 
\item (iii) generic pNGBs arising from the spontaneous breaking of a new U(1) global symmetry at some large energy scale $\Lambda$, like \eg\ the familon (pNGB of a flavour-chiral U(1) horizontal symmetry invoked to explain the fermion families)~\cite{Davidson:1981zd,Wilczek:1982rv}, the R-axion (pNGB of the broken R-symmetry in SUSY)~\cite{Bellazzini:2017neg}, pseudoscalar states in Higgs compositeness models (from the broken U(1) symmetry acting on all underlying fermions of the theory)~\cite{Cacciapaglia:2019bqz}, or in cosmic inflation~\cite{Freese:1990rb}; 
\item (iv) generic pseudoscalar bosons that appear \eg\ in extended Higgs sectors~\cite{Branco:2011iw}, 
in phenomenological realizations of string theory~\cite{Ringwald:2012cu}, in models where they contribute to lepton dipole moments (thereby solving the $(g-2)_\mu$ problem)~\cite{Chang:2000ii,Marciano:2016yhf} or address the electroweak hierarchy problem~\cite{Graham:2015cka}.
\end{description}

\begin{figure}[htbp!]
\centering
\includegraphics[width=0.65\columnwidth]{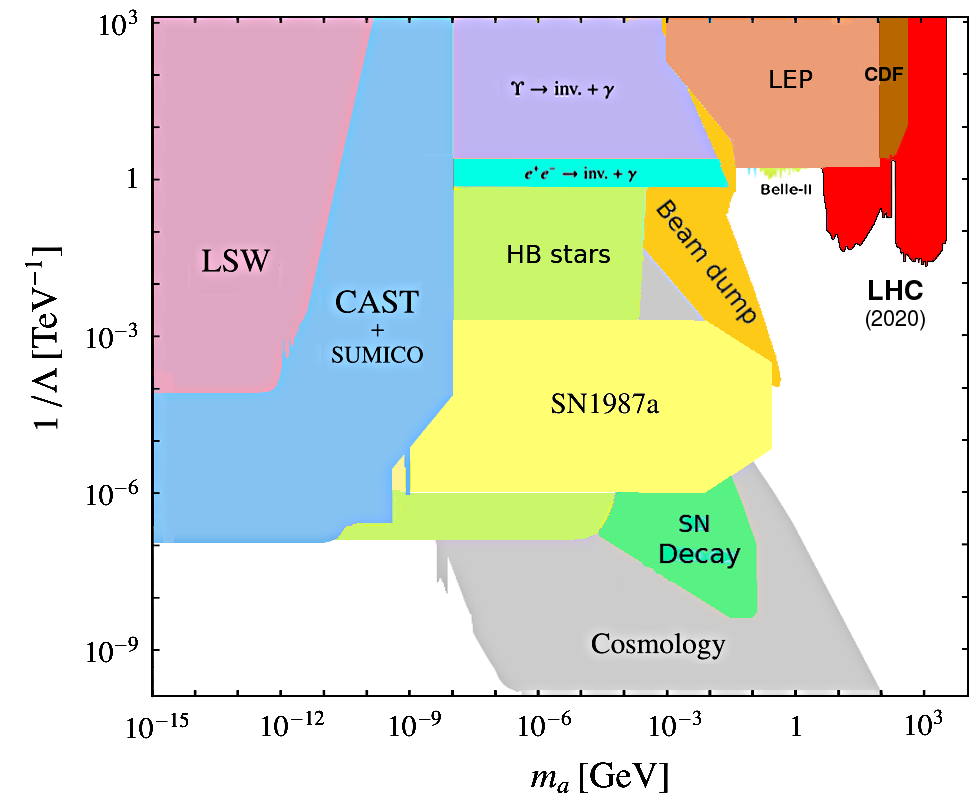}
\caption{Limits on the ALP-photon coupling vs.\ ALP mass from cosmological, astrophysical, and accelerator searches. The top-right red contour depicts the approximate area currently covered by the latest LHC studies summarized here. Figure adapted from~\cite{Jaeckel:2015jla,Bauer:2017ris}.}
\label{fig:limits_preLHC}
\end{figure}

Since light pseudoscalars naturally couple to photons (due to spin selection rules and mass\footnote{For a very light ALP with $m_a < 2m_e$, the diphoton channel is the only SM decay mode allowed.} constraints), ALP searches from cosmology, astrophysical, and low-energy accelerator studies have mostly exploited the (inverse) Primakoff effect~\cite{Halprin:1966zz}, $\gaga^{(*)}\to a \to \gaga^{(*)}$, that converts photons into axions (and vice versa) in collisions with other photons or, equivalently, electromagnetic fields ($\gamma^*$). Collider searches for ALPs have thereby predominantly focused on initial and/or final states with photons that, first, allow a direct connection with the vast range of all other existing searches and, secondly, are easier to identify over intrinsically large hadronic backgrounds.
Focusing on the ALP-$\gamma$ coupling, the effective Lagrangian reads:
\begin{equation}
\mathcal{L} \supset -\frac{1}{4}g_{a\gamma}\,a\,F^{\mu\nu}\Tilde{F}_{\mu\nu},\, \mbox{ with } g_{a\gamma} = C_{\gaga}/\Lambda,
\label{eq:L_agamma}
\end{equation}
where $a$ is the ALP field, $F^{\mu\nu}$ ($\Tilde{F}_{\mu\nu}$) is the photon field strength (dual) tensor, and the dimensionful ALP-photon coupling strength $g_{a\gamma}$ is related\footnote{For the QCD axion, decay constant and mass are directly related and $g_{a\gamma}$ is proportional to $m_a$ with an essentially known fixed constant
(thereby populating a linear band in Fig.~\ref{fig:limits_preLHC}), but such a relationship is relaxed for generic ALPs, which do not necessarily couple to gluons, and for which $m_a$ and $g_{a\gamma}$ are free parameters.} 
to the high-energy scale $\Lambda$ associated with the broken symmetry in the ultraviolet (the effective coefficient $C_{\gaga}$ rescales appropriately the $a$-$\gamma$ coupling whenever the ALP couples to other SM particles, most often $C_{\gaga}=1$ is considered hereafter). 
The whole ALP phenomenology (production, decay, interaction with magnetic fields) is thereby fully defined in the $(m_a,\,g_{a\gamma})$ parameter space.

Figure~\ref{fig:limits_preLHC} shows the current limits over about 19 orders of magnitude in ALP mass and 13 orders in ALP-$\gamma$ coupling.
In the very light mass range ($m_a\lesssim 10$~eV), the most stringent limits are provided by light-shining-through-a-wall (LSW) experiments, as well as solar photon instruments like the Tokyo Axion Helioscope (SUMICO) and the CERN Axion Solar Telescope (CAST)~\cite{Graham:2015ouw}. 
A combination of cosmological constraints (number of effective degrees of freedom, primordial big-bang nucleosynthesis, distortions of the cosmic microwave-background spectrum, diffuse extragalactic photon measurements) leads to the grey region at the lowest $g_{a\gamma}$ couplings, over $m_a \approx 1$~keV--10~GeV~\cite{Cadamuro:2011fd,Millea:2015qra} (recent works~\cite{Depta:2020wmr} have, however, weakened the cosmological constraints in the ``triangle'' region 
between the beam dump and SN1987a bounds). For ``intermediate'' masses and couplings, astrophysical bounds have been derived from energy loss of stars
through ALP radiation, which are constrained by the ratio of red giants to younger stars in the so-called horizontal branch (HB, light green), the measurement of the length of the neutrino burst from Supernova SN1987a (yellow)~\cite{Chang:2018rso}, and the absence of SN photon bursts from decays of produced ALPs (green).

Experimental attention to ALPs above the MeV scale at particle accelerators is relatively recent. Since the $a \to \gaga$ decay rate scales with the third power of the ALP mass, for decreasing values of $m_a$ the decay rate becomes so small that the ALP leaves the detector and appears as an invisible particle. In fixed-target proton and electron experiments (orange ``beam dump'' band), photons copiously produced via meson decays, or via bremsstrahlung off electron beams, can convert into ALPs via the Primakoff process off a nuclear target, and ALPs are searched for in final states with single photons (inverse Primakoff scattering process) or diphotons (from $a\to\gaga$ decays) emitted in the forward direction~\cite{Dobrich:2015jyk,Banerjee:2020fue}. The top area of Fig.~\ref{fig:limits_preLHC} is dominated by searches at $\epem$ colliders via mono-photon final states with missing energy from long-lived ``invisible''\footnote{Such final states include also the case where the ALP is a short-lived dark-matter mediator that decays into invisible DM particles.} ALPs via radiative $\Upsilon\to\gamma a$ decays (violet area) and $\epem\to \gamma a$ (light blue area) at CLEO and BaBar~\cite{Balest:1994ch,delAmoSanchez:2010ac}, and diphoton and triphoton final states ($\epem\to 2\gamma, 3\gamma$), for $m_a=50$~MeV--8~GeV and $m_a=20$--100~GeV respectively, at LEP-I and II~\cite{Acciarri:1994gb,Abbiendi:2002je} (light brown) derived in~\cite{Jaeckel:2015jla,Knapen:2016moh}, as well as similar searches at Belle-II~\cite{BelleII:2020fag}, PrimEx~\cite{Aloni:2019ruo}, and CDF~\cite{Aaltonen:2013mfa}. As one can see from Fig.~\ref{fig:limits_preLHC}, before the LHC, ALP limits for masses above $m_a \approx 1$~GeV were very scarce (and nonexistent above $m_a\approx 300$~GeV), and this is the region where the energy-frontier collider plays a unique role. 

\begin{figure}[htbp!]
\centering
\includegraphics[width=0.7\columnwidth]{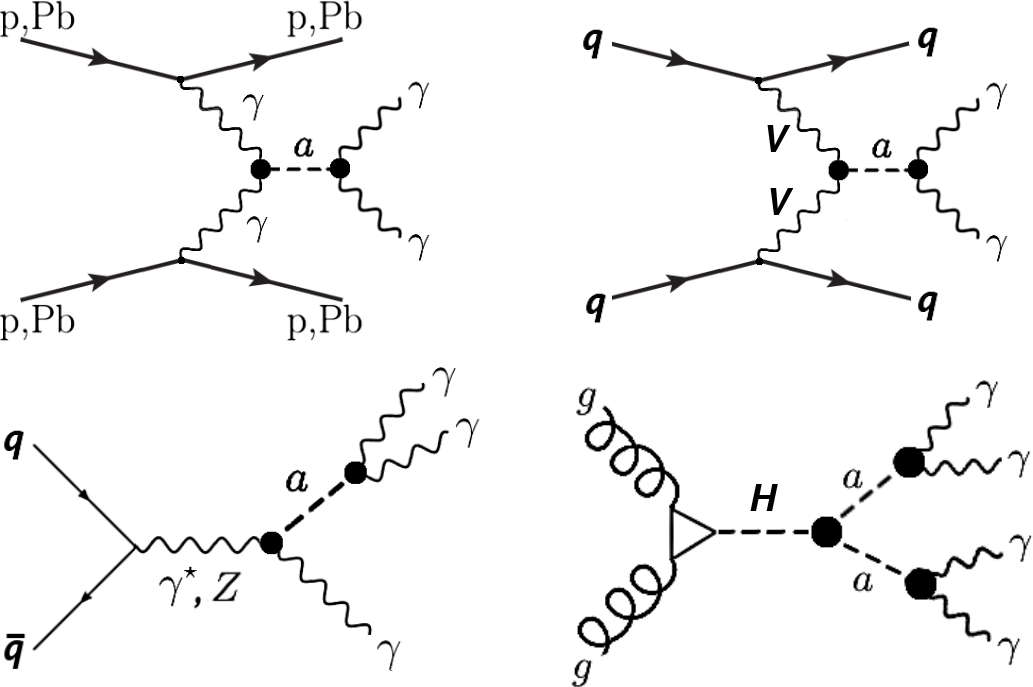}
\caption{Representative diagrams of ALPs searches at the LHC via exclusive diphotons (top left), diphotons from vector-boson-fusion (with $V = \gamma, \mathrm{Z}$, top right), exotic triphoton Z-boson decays (bottom left), and exotic 4-photon Higgs-boson decays (bottom right).}
\label{fig:diags}
\end{figure}

Searches for axion-like particles 
at the LHC share the following properties:
\begin{description}
\item (1) Final states with photons plus, in some cases, Z and H bosons, are considered;
\item (2) if ALPs are long-lived (usually for $m_a \lesssim1$~GeV), bounds can be derived from monophoton final-states with missing transverse energy (from the escaping $a$) in the LHC detectors;
\item (3) if ALPs are short-lived (usually for $m_a \gtrsim 1$~GeV), bounds are set from their decays into photons inside the LHC detectors volume.
\end{description}
For case (2) above, LHC limits on ALPs with masses below 1~GeV in monophoton topologies have been discussed in detail in~\cite{Mimasu:2014nea} and, since they basically overlap with the beam-dump bounds shown in Fig.~\ref{fig:limits_preLHC}, will not be discussed further. Instead, for case (3) above, competitive searches have been set for masses $m_a > 5$~GeV in 2-, 3-, 4-photon final states via searches for exclusive and inclusive $\gaga$ resonances, and exotic Z or Higgs boson decays, with the corresponding diagrams shown in Fig.~\ref{fig:diags}. 
The ALP bounds have been set in some cases directly by the LHC experiments themselves, but mostly by phenomenological reinterpretations of experimental data from ATLAS~\cite{Aad:2008zzm} and CMS~\cite{Chatrchyan:2008aa} based on generic spin-0 $\gaga$ resonance searches, as discussed case-by-case below.

\section{ALP searches in exclusive $\gaga$ final states}

Proton (p) and lead ions (Pb) accelerated at LHC energies generate strong electromagnetic (e.m.) fields that, in the equivalent photon approximation~\cite{Budnev:1974de}, can be identified as quasireal photon fluxes with very low virtualities $Q^{2} < 1/R_A^{2}\approx 0.08,\,8\cdot10^{-4}$~GeV$^2$ and longitudinal energies as large as $\omega_{\rm max} = \gamma_L/R_A \approx 2.5$~TeV, 80~GeV, respectively\footnote{Taking $R_A\approx 0.7,7$~fm as p, Pb charge radii, and $\gamma_L=E_{\rm beam}/m_{p,N}$ for their beam Lorentz factors.}~\cite{Baltz:2007kq,dEnterria:2013zqi}. On the one hand, in the Pb-Pb case, since the photon flux scales as the square of the charge of each colliding particle, $\gaga$ cross sections are enhanced by up to a factor of $Z^4 \approx 5\cdot 10^{7}$ compared to p-p (or $\epem$) collisions, although they reach relatively moderate maximum center-of-mass (\cm) energies of the order of $\sqrt{s^{\rm max}_{\gaga}}\approx 200$~GeV. On the other hand, photon-photon processes in p-p collisions, although featuring much lower $\gamma$ fluxes (compensated, in part, by the much larger beam luminosities compared to heavy ions), can reach much higher $\sqrt{s^{\rm max}_{\gaga}}\approx 4.5$~TeV values, thanks to the harder proton $\gamma$ fluxes. Exploiting such interesting photon-photon possibilities to search for axions in the so-called ``ultraperipheral'' collisions at the LHC via the diagram shown in Fig.~\ref{fig:diags} (top left) was suggested in~\cite{Knapen:2016moh,Knapen:2017ebd}. The final state of interest is that of an exclusive diphoton event, \ie\ without any other particle produced, with the two photons emitted back-to-back (the ALP is produced basically at rest due to the low virtuality of the colliding photons) and featuring a peak on top of the invariant mass ($m_{\gaga}$) distribution of the light-by-light (LbL) scattering $\gaga\to\gaga$ continuum~\cite{dEnterria:2013zqi}. In p-p collisions, the presence of very large pileup events hinders the observation of this process unless one can tag one or both protons in very forward spectrometers, such as the CMS-TOTEM PPS and ATLAS AFP ones~\cite{Royon:2015tfa}, to remove overwhelming hadronic backgrounds. The absence of pileup in Pb-Pb collisions makes this colliding mode the most competitive one for exclusive ALP searches up to $\sqrt{s^{\rm max}_{\gaga}}\approx 100\,(300)$~GeV compared to p-p running with proton taggers located at 420 (220)~m from the interaction point (IP)~\cite{Bruce:2018yzs}.

The first exclusive diphoton search at the LHC was carried out by CMS in p-p collisions at 7~TeV with the first 36-pb$^{-1}$ of integrated luminosity~\cite{Chatrchyan:2012tv}. The event selection consisted of two photons with transverse energy $\ET>2$~GeV and pseudorapidity $|\eta|<2.5$ with no other hadronic activity over $|\eta|<5.2$. The lack of observed events imposes an upper limit cross section of $\sigma(\pp\to\rm{p}\gaga\rm{p})>1.18$~pb at 95\% confidence level (CL). Such a result was subsequently recast into the $(m_a,g_{a\gamma})$ plane (red area in Fig.~\ref{fig:limits_VBF}, right)~\cite{Knapen:2016moh,Knapen:2017ebd}, overlapping with the LEP-II limits over $m_a \approx 5$--90~GeV. A recent search of the same exclusive final state has been carried out by CMS\,+\,TOTEM in p-p collisions at 13~TeV with forward proton tagging, requiring two photons with $\ET>75$~GeV over $|\eta|<2.5$, with $m_{\gaga}> 350$~GeV and low acoplanarity~\cite{CMS:2020rzi}. Matching the diphoton mass and pseudorapidity measured in the central detectors with those of the CT-PPS proton spectrometers allows the removal of p-p pileup events. No exclusive $\gaga$ event is found above expected backgrounds, leading to an upper limit cross section of $\sigma(\pp\to\rm{p}\gaga\rm{p})>3.0$~fb at 95\% CL, that has been recast into limits on anomalous quartic photon couplings over $m_{\gaga}\approx 0.4$--2~TeV~\cite{CMS:2020rzi}. However, in terms of ALP bounds in this mass range, such a measurement is not yet competitive compared to other $\gaga$ final states discussed below.\\

\begin{figure}[htbp!]
\centering
\includegraphics[width=0.48\columnwidth,height=5.5cm]{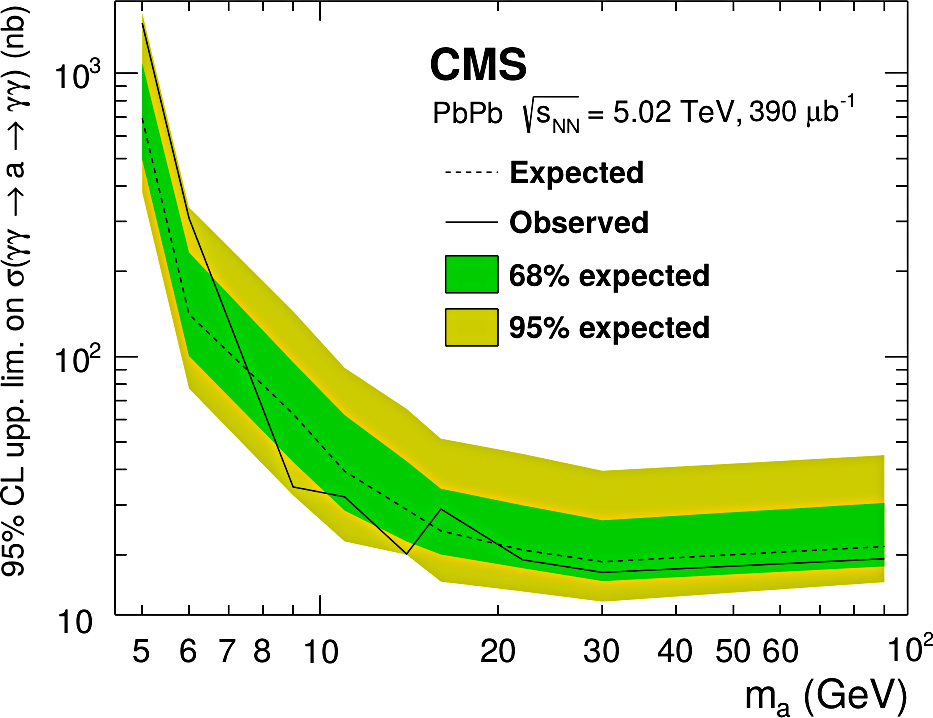}\hspace{0.5cm}
\includegraphics[width=0.48\columnwidth]{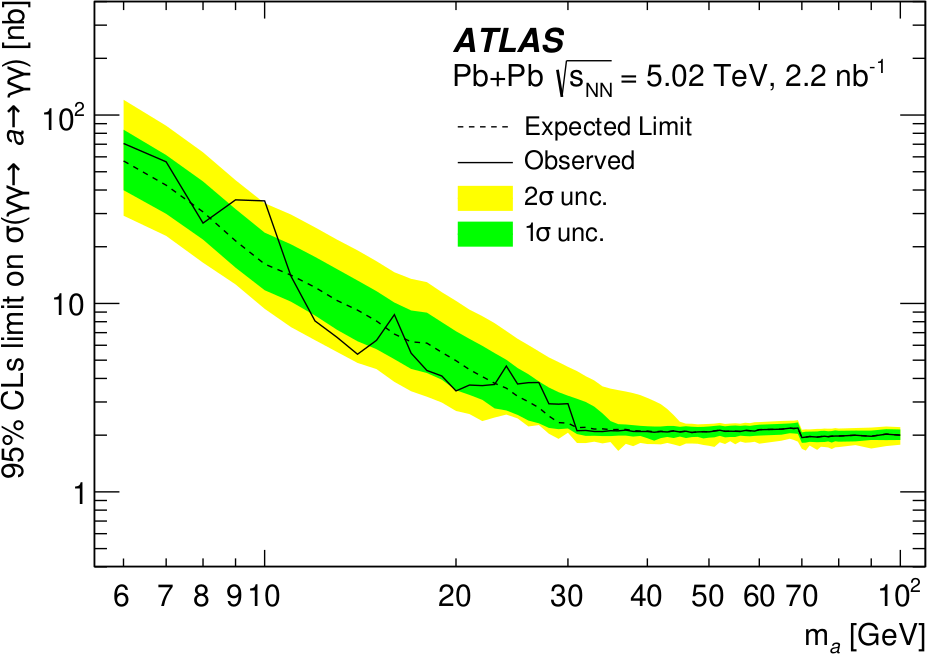}
\caption{95\% CL upper limits on the ALP cross section $\sigma(\gaga\to a\to\gaga)$ as a function of ALP mass $m_a$ measured in ultraperipheral Pb-Pb collisions at $\sqrtsnn = 5.02$~TeV in CMS (left)~\cite{Sirunyan:2018fhl} and ATLAS (right)~\cite{Aad:2020cje}. The observed (expected) upper limits are shown as solid (dashed) black curves (with $\pm1$ and $\pm2$ standard deviation bands).}
\label{fig:sigma_UPC_ALPs}
\end{figure}

The most stringent limits on ALPs over $m_a\approx 5$--100~GeV have been set by searches in ultraperipheral Pb-Pb collisions by CMS~\cite{Sirunyan:2018fhl} and ATLAS~\cite{Aad:2020cje}. The final state of interest corresponds to two photons with $\ET>2,3$~GeV, $|\eta|<2.5$, $m_{\gaga}>5$~GeV with no other hadronic activity over $|\eta|<5$. Extra photon-pair kinematic criteria are applied to remove any remaining background (mostly from misidentified $\gaga\to\epem(\gamma)$ events): $\pT^{\gaga}<1$~GeV, and acoplanarity $\Delta\phi_{\gaga}<0.01,0.03$, which enhance the photon-fusion production characterized by two back-to-back photons. Both ATLAS and CMS observe a number of exclusive diphoton counts consistent with the expectations of the very rare LbL scattering process~\cite{dEnterria:2013zqi}. For ALP searches, both experiments have subsequently injected into their measured $m_{\gaga}$ distribution, simulated $\gaga\to a\to\gaga$ peaks generated with \starlight~\cite{Klein:2016yzr} with the ALP-$\gamma$ coupling implemented as described in~\cite{Knapen:2016moh}.
The diphoton trigger and reconstruction efficiencies are found to be $\sim$20\% (50\%) at $m_a=5$ (50)~GeV, with the assumed ALP width dominated by the experimental $\gaga$ mass resolution. 
It is worth to note that such experimental effects are not always taken appropriately into account by phenomenological works, which often use very simplified detector inefficiencies, when reinterpreting existing data to extract new ALP limits.\\
The absence of any visible peak above the LbL continuum is used to set the expected and observed limits on the ALPs production cross section as a function of mass shown in Fig.~\ref{fig:sigma_UPC_ALPs}. The ATLAS results (right plot) exploit the largest integrated luminosity of the 2018 Pb-Pb run, and allow one to exclude ALP integrated cross sections above 2 to 70~nb at the 95\% CL over the $m_a = 6$--100 GeV range. The cross section limit from CMS (left plot) provides, in the bin $m_a = 5$--6~GeV, the lowest ALP mass probed so far at the LHC. It is unlikely that ATLAS or CMS will be able to go to lighter masses due to the difficulties in triggering and reconstructing very low $\ET\lesssim 2$~GeV photons. However, LHCb and ALICE have good photon reconstruction capabilities down to $\sim$1~GeV (or lower, exploiting $\gamma\to\epem$ conversions) and could improve those limits in a mass region $m_a \approx 0.5$--5~GeV of otherwise difficult experimental access, thereby bridging with the beam-dump and BELLE-II constraints in the ``wedge'' area shown in Fig.~\ref{fig:limits_preLHC}. Experimental challenges in this mass region are the $m_{\gaga}$ resolution, and the presence of multiple spin-0 and 2 resonances with visible diphoton decays~\cite{Klusek-Gawenda:2019ijn}.

The limits on the ALP cross section (with signal strength $\sqrt{\mu_{\rm CLs}}$) are transformed into the corresponding limits for the ALP-$\gamma$ coupling $g_{a\gamma}\equiv 1/\Lambda$, assuming a BR$(a\to\gaga)=100\%$ branching ratio, \ie\ effectively setting $C_{\gaga}=1$ in Eq.~(\ref{eq:L_agamma}), calculated from $1/\Lambda = \sqrt{\mu_{\rm CLs}}\cdot (1/\Lambda^{\rm MC})$ where the latter is the coupling used in the MC generator. The derived $1/\Lambda$ constraints range from $g_{a\gamma}\approx 1$~TeV$^{-1}$ to 0.05~TeV$^{-1}$ (violet area in Fig.~\ref{fig:limits_final}, right). It is worth to mention that the CMS Pb-Pb ALP results have been also recast into limits on the $(m_a,g_{aB})$ plane, including also the hypercharge coupling $g_{aB}$, \ie\ processes involving the Z boson discussed in Section~\ref{sec:gammaZ}. Locally, these bounds are the strongest ones over $m_a\approx 25$--60~GeV~\cite{Sirunyan:2018fhl}, above the $\pp\to 3\gamma$ ones discussed later.

\section{ALP searches in generic $\gaga$ final states}

In p-p collisions at the LHC, ALPs can also be produced in parton-parton scatterings, via gluon-fusion and vector-boson-fusion (VBF) processes, followed by their diphoton decay. Since one usually focuses on ALP couplings to electroweak bosons, and in particular to photons in order to compare them to existing cosmological and astrophysical limits, the gluon-fusion production mode will not be further considered here\footnote{Detailed discussions on exploiting the LHC data to place bounds on ALPs sensitive to both photons and gluons can be found \eg\ in~\cite{Jaeckel:2012yz,Mimasu:2014nea,Mariotti:2017vtv,CidVidal:2018blh,Gavela:2019cmq}.} and we will only discuss diphoton final states from VBF processes such as the one shown in Fig.~\ref{fig:diags} (top-right).

\begin{figure}[htbp!]
\centering
\includegraphics[width=0.45\columnwidth,height=5.5cm]{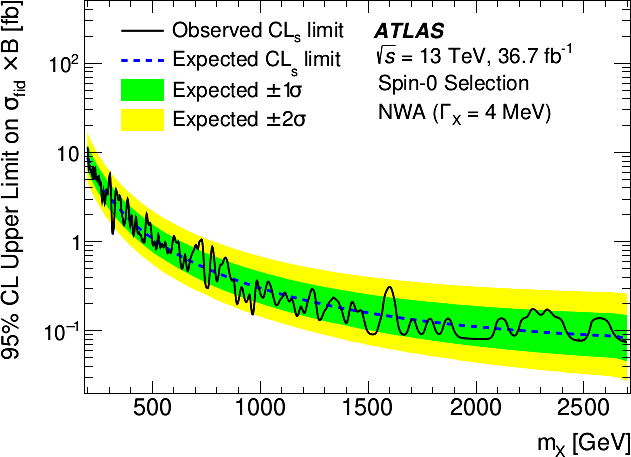}\hspace{0.1cm}
\includegraphics[width=0.51\columnwidth,height=5.7cm]{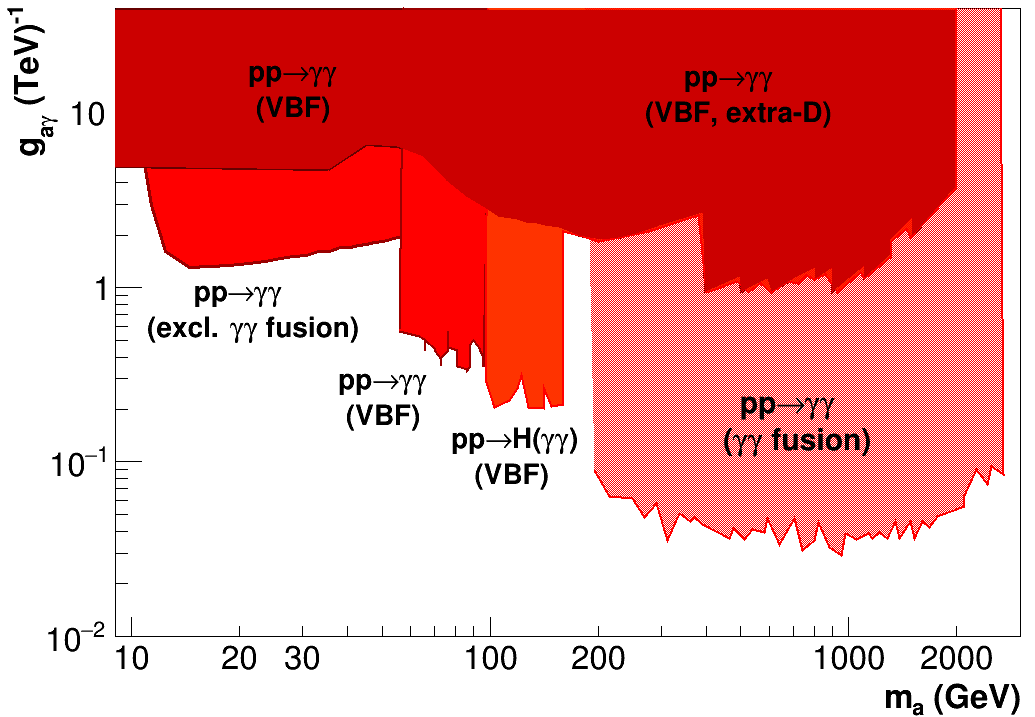}
\caption{Left: Upper limits on the fiducial cross section times diphoton branching ratio of a narrow-width ($\Gamma_X = 4$~MeV) spin-0 resonance as a function of its mass $m_X$ in p-p collisions at $\sqrts = 13$~TeV~\cite{Aaboud:2017yyg}.
Right: Derived exclusion regions on the ALP-$\gamma$ coupling vs.\ ALP mass plane from various diphoton production data sets in p-p collisions at the LHC~\cite{Jaeckel:2012yz,Jaeckel:2015jla,Knapen:2016moh,Bauer:2018uxu}.}
\label{fig:limits_VBF}
\end{figure}

Limits on the production of ALPs have been derived in~\cite{Jaeckel:2012yz,Jaeckel:2015jla,Knapen:2016moh,Bauer:2018uxu} based on reinterpretations of existing searches of Higgs, generic spin-0, and/or extra dimensions, as well as QCD studies, in diphoton final states measured by ATLAS and CMS over the range $m_{\gaga}\approx 10$~GeV--2.6~TeV in p-p collisions at $\sqrts = 8, 7, 13$~TeV (Fig.~\ref{fig:limits_VBF}, right).
First, the ATLAS QCD diphoton data at 7~TeV~\cite{Aad:2012tba} has been used to set limits on the ALP-photon coupling of the order of $g_{a\gamma}\approx 5$~TeV$^{-1}$ over $m_a = 10$--100~GeV.
The ALP diagram of Fig.~\ref{fig:diags} (top-right) is identical to Higgs boson production via VBF, accompanied by forward and backward jets from the radiating quarks, followed by its diphoton decay, with kinematic differences between scalar and pseudoscalar resonances being of the order of $\mathcal{O}$(10\%)~\cite{Jaeckel:2012yz}. Therefore, in the region $m_a = 100$--160~GeV, the VBF data from the ATLAS Higgs observation at 7 and 8~TeV~\cite{Aad:2012tfa}, establishing the maximum allowed cross sections in each mass bin, has been exploited to set the corresponding ALP bounds (orange band in Fig.~\ref{fig:limits_VBF}). For $m_a = 150$--400~GeV and $m_a = 400$--2000~GeV, the CMS~\cite{Chatrchyan:2011jx,Chatrchyan:2011fq} and ATLAS~\cite{ATLAS:2011ab} extra-dimension searches at $\sqrts = 7$~TeV respectively, have been used. 

In addition, the work of~\cite{Knapen:2016moh} exploited the ATLAS searches for scalar diphoton resonances at $\sqrts =8$~TeV~\cite{Aad:2014ioa} to set limits over $m_{a} =65$--100~GeV (red rectangular region in Fig.~\ref{fig:limits_VBF}, right), which are basically coincident with those obtained from the OPAL $\epem\to 3\gamma$ studies at LEP~\cite{Abbiendi:2002je}. More recently, Ref.~\cite{Bauer:2018uxu} has computed the constraints on $a\to\gaga$ produced via photon-fusion in p-p collisions, employing the photon distribution functions in the proton from Ref.~\cite{Manohar:2016nzj}, to reinterpret the ATLAS upper limits on the cross sections for spin-0 resonances in 39.6 fb$^{-1}$ of p-p data collected at 13~TeV~\cite{Aaboud:2017yyg} (Fig.~\ref{fig:limits_VBF}, left). This work leads to the most competitive limits today over $m_{a} = 200$--2600~GeV (lighter red area in Fig.~\ref{fig:limits_VBF}, right). It would be appropriate, however, for LHC experiments to revisit these latter limits, as their derivation may be too optimistic in the treatment of the relevant backgrounds~\cite{Florez:2021zoo}. 


\section{ALP searches in exotic Z and H boson photon decays}
\label{sec:gammaZ}

\begin{figure}[htbp!]
\centering
\includegraphics[width=0.41\columnwidth]{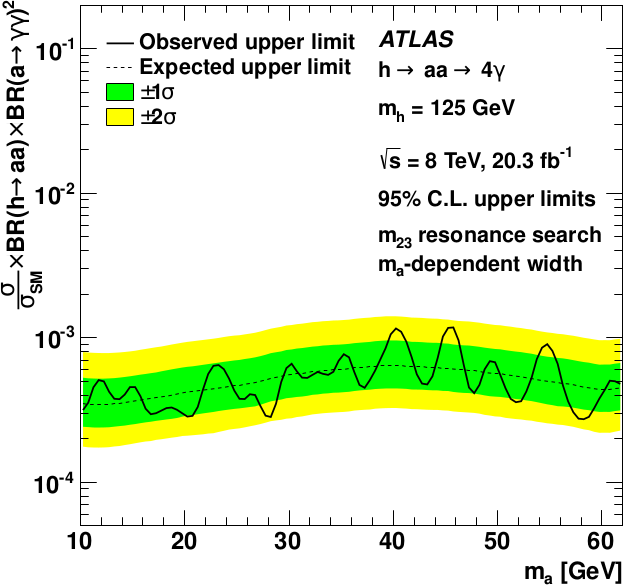}
\hspace{0.1cm}
\includegraphics[width=0.56\columnwidth]{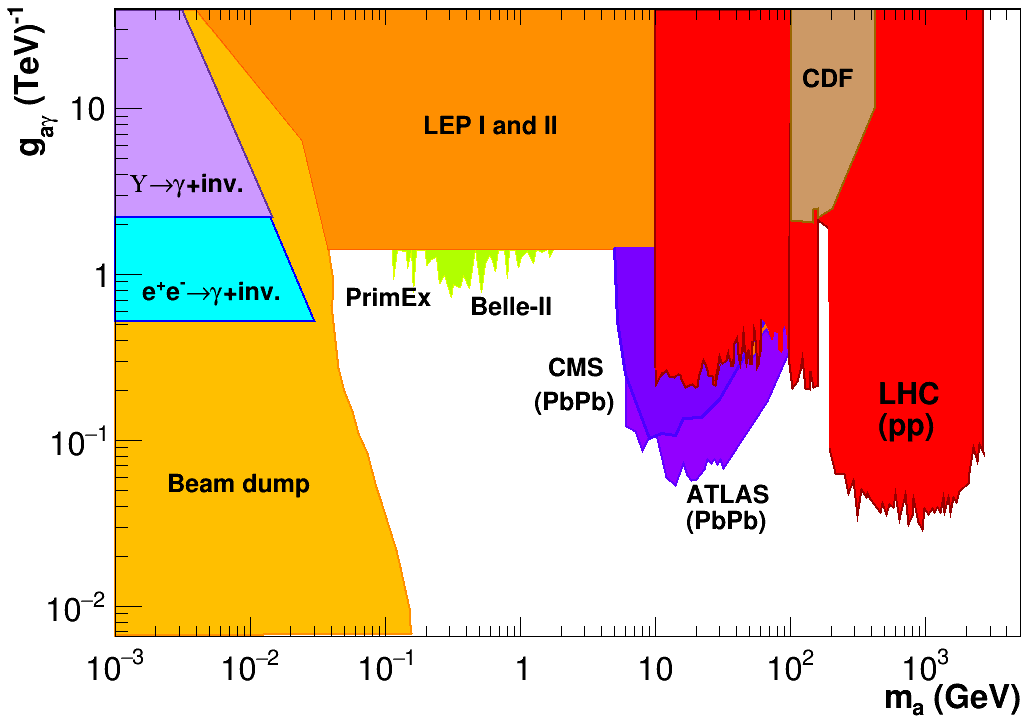}
\caption{Left: 95\% CL upper limits on $(\sigma/\sigma_{\rm SM}) \times$\,BR$(\mathrm{H}\to aa) \times$ BR$(a \to\gaga)$ with $\pm 1\sigma$ and $\pm2\sigma$ uncertainty bands from the resonance search hypothesis tests, accounting for statistical and systematic uncertainties from simulated signal samples~\cite{Aad:2015bua}.
Right: Detailed bounds in the $(m_a,g_{a\gamma})$ plane from all existing accelerator and collider ALP searches for masses $m_a\approx 1$~MeV--3~TeV. The LHC constraints summarized in this work are indicated by the area in red (p-p) and violet (Pb-Pb collisions).}
\label{fig:limits_final}
\end{figure}

If one considers the possibility that the ALP couples not only to photons, but also to hypercharge bosons $B$, one can generalize Eq.~(\ref{eq:L_agamma}) to the following effective Lagrangian
\begin{equation}
\mathcal{L} \supset 
-\frac{1}{4}g_{aB}\,a\,B^{\mu\nu}\Tilde{B}_{\mu\nu} \,.
\label{eq:L_agammaZ}
\end{equation}
The individual ALP-electroweak couplings 
can be related to each other through the underlying effective Wilson coefficients $C_{\gaga} = C_{\rm WW} + C_{BB}$,  $C_{\gamma \mathrm{Z}} = c_{\theta}^2\,C_{\rm WW} -s_{\theta}^2\,C_{BB}$, and $C_{\mathrm{Z}\mathrm{Z}} = c_{\theta}^4\,C_{\rm WW}+s_{\theta}^4\,C_{BB}$, with $s_{\theta}$ ($c_{\theta}$) being the (co)sine of the weak mixing angle~\cite{Dolan:2017osp,Bauer:2018uxu}. The exotic decay $\mathrm{Z} \to \gamma a$ (Fig.~\ref{fig:diags}, bottom left)  is governed by the coefficient $C_{\gamma \mathrm{Z}}$. When interpreting the corresponding Z-boson experimental data in the $(m_a,g_{a\gamma})$ plane, one often assumes that the two effective coefficients $C_{\gaga}$ and $C_{\gamma \mathrm{Z}}$ are correlated, \eg\ if the ALP couples to hypercharge but not to SU(2)$_L$, then $C_{\gamma \mathrm{Z}} = -s_{\theta}^2 C_{\gaga}$ since $C_{\rm WW} = 0$. The search for unusual triphoton decays of onshell Z bosons at colliders thereby provides sensitivity to ALPs with masses $m_a\lesssim m_\mathrm{Z} \approx 90$~GeV~\cite{Jaeckel:2015jla,Brivio:2017ije,Bauer:2017ris}. The LEP bounds shown in Fig.~\ref{fig:limits_preLHC} include ALP searches via triphoton final states on and off the Z pole at LEP~\cite{Jaeckel:2015jla,Abbiendi:2002je} and in $\ppbar$ collisions at Tevatron (CDF)~\cite{Aaltonen:2013mfa} (brown area), assuming $C_{\rm WW} = 0$. Given the much larger number of Z bosons measured at the LHC, significant improvements in such limits are expected from the corresponding ATLAS and CMS data sets. 
The most stringent constraint is set by the ATLAS measurement of BR$(\mathrm{Z} \to 3\gamma) < 2.2\cdot 10^{-6}$ at 95\% CL~\cite{Aad:2015bua}, improving by a factor of five the previous BR$<10^{-5}$ result at LEP~\cite{Acciarri:1994gb}. Different phenomenological analyses~\cite{Knapen:2016moh,Bauer:2017ris} have recast such a result into limits in the $(m_a,g_{a\gamma})$ plane, covering the region $m_a \approx 10$--70~GeV, the latter value determined from the experimental requirement of $\pT^{\gamma} > 17$~GeV (red area below the Pb-Pb bounds in Fig.~\ref{fig:limits_final} right). It is worth noting that the phenomenological recast of the $\pp\to 3\gamma$ data is subject to intrinsic 3-photon combinatorics uncertainties, and that ideally such ALP limits extraction should be redone by the experiments themselves.

Searches for decays of the Higgs boson to a pair of pseudoscalar bosons  (Fig.~\ref{fig:diags}, bottom-right) have been also a common BSM search channel at the LHC in the last years, motivated \eg\ by Higgs compositeness~\cite{Cacciapaglia:2019bqz} and extended Higgs sectors~\cite{Branco:2011iw} models. Whereas the Lagrangian (\ref{eq:L_agammaZ}) encoding the coupling of ALPs to electroweak gauge bosons is of dimension-5, interactions of the ALP with the Higgs boson appear only at dimension-6 and higher, with two operators mediating the decay $\mathrm{H}\to aa$ and $\mathrm{H}\to \mathrm{Z} a$ respectively~\cite{Bauer:2017ris}. Although searches for the exotic Higgs decays $\pp \to \mathrm{H}\to \mathrm{Z} a \to \mathrm{Z} \gaga$ and $\pp \to \mathrm{H}\to aa \to 4\gamma$ cannot be directly translated into constraints in the $(m_a,g_{a\gamma})$ plane, because the corresponding ALP-Higgs coefficients $C_{H\mathrm{Z}}$ and $C_{Ha}$ are generally not related to $C_{\gaga}$, these processes probe the parameter space in which an ALP can provide the explanation of the anomalous magnetic moment of the muon~\cite{Bauer:2017ris}. The ATLAS collaboration has carried out a search of the $\mathrm{H}\to aa$ decay in p-p collisions at 8~TeV~\cite{Aad:2015bua} (Fig.~\ref{fig:limits_final}, left), which has been subsequently translated into exclusion limits on the $(m_a,g_\mathrm{aH})$ plane at low ($m_a<400$~MeV) and high ($m_a=10$--62.5~GeV) ALP masses~\cite{Bauer:2017ris,Bauer:2018uxu}. 


\section{Summary and outlook}

High-energy colliders provide unique possibilities to explore heavy axion-like particles (ALPs), with masses above $m_a\gtrsim 1$~GeV, 
that otherwise lie beyond the reach of lower energy accelerators and of existing cosmological and astrophysical constraints. ALPs with masses above $m_a\approx 1$~GeV can only be accessed via collider experiments (except for extremely weakly coupled ones, with $g_{a\gamma}<10^{-6}$~TeV$^{-1}$, excluded by cosmological observations), with the LEP and Tevatron data providing the best limits up to $m_a\approx 100,\,300$~GeV before the LHC started operation 10 years ago. Searches at the CERN LHC for ALPs decaying into two photons, $a\to\gaga$, have been summarized, focusing on scenarii with dominant $a$-$\gamma$ couplings. Different channels have been scrutrinized including exclusive and inclusive diphoton production in proton-proton and lead-lead collisions, $\pp,\,\PbPb \to a \to \gaga\,(+X)$, as well as exotic Z and Higgs boson decays, $\pp \to \mathrm{Z},\mathrm{H}\to a\gamma \to 3\gamma$ and $\pp \to \mathrm{H}\to aa \to 4\gamma$. The limits in the ALP-photon coupling vs.\ ALP mass plane $(m_a,g_{a\gamma})$ derived from the LHC data extend by one-order-of-magnitude the range of preceding bounds in both mass (from $m_a\approx 400$~GeV previously at Tevatron, up to 2.6~TeV now) and in $g_{a\gamma}$ for masses $m_a\approx 100$~GeV (from $g_{a\gamma} = 0.5$~TeV$^{-1}$ previously at LEP, down to 0.05~TeV$^{-1}$ now). 
Exclusive diphoton searches in PbPb collisions provide now the best ALP exclusion~limits for masses $m_a\approx 5$--100~GeV, whereas the other channels are the most competitive ones over $m_a\approx 100$~GeV--2.6~TeV.
ALP searches in exotic Z boson decays have been also exploited to set bounds in the $m_a\approx 10$--70~GeV range, with specific assumptions about the relationship between the $C_{\gaga}$ and $C_{\gamma \mathrm{Z}}$ coupling operators. Searches for exotic 3- and 4-photon Higgs boson decays through intermediate ALPs provide also strong constraints on new dimension-6 and -7 operators. Although the latter cannot be recast into bounds in the standard $(m_a,g_{a\gamma})$ plane, the study of several couplings simultaneously is crucial to identify the most interesting regions in ALP parameter space.

The LHC constraints summarized here are indicated by the red (for p-p collisions) and violet (for Pb-Pb collisions) areas in Fig.~\ref{fig:limits_final} (right), compared to previous fixed-target and collider results. In the next 15 years, the currently uncovered (white) areas in this plot will be probed by many different experiments. The now ``empty'' region $m_a\approx 50$~MeV--5~GeV will be accessible to various future experiments both descending along the vertical $g_{a\gamma}$ axis alone, such as Belle-II  with 50~ab$^{-1}$ (improving its current limits at $g_{a\gamma}\approx 1$~TeV$^{-1}$ by two orders of magnitude, over $m_a\approx 0.2$--8~GeV)~\cite{Dolan:2017osp} and GlueX (50 times better than the local PrimEx limit today of $g_{a\gamma}\approx 1$~TeV$^{-1}$, over $m_a\approx 50$--500~MeV)~\cite{Aloni:2019ruo}, or extending the beam-dump ``wedge'' in both $g_{a\gamma}$ and $m_a$ directions at new CERN experiments searching for long-lived particles such as \eg\  SHiP~\cite{Alekhin:2015byh},
FASER~\cite{Feng:2018pew}, and MATHUSLA~\cite{Alpigiani:2018fgd}.

The full completion of the LHC physics program will provide about 10 and 100 times more integrated luminosities in heavy-ion and p-p collisions, respectively, than analyzed in ALP searches so far. The clean searches via exclusive $\gaga\to a\to\gaga$ final states will benefit, in particular, from the possibility of forward proton-tagging (as well as new ion species, including p-Pb running) that will further extend the LHC coverage in mass regions $m_a\approx 2.6$--5~TeV (and $m_a\approx 200$--400~GeV) with non-existent (or locally weaker) bounds today~\cite{Baldenegro:2018hng,Schoeffel:2020svx}. Thus, ultimate 10--100 improvements in the ALPs bounds are expected at the LHC with ALP-photon couplings approaching $g_{a\gamma}\approx 10^{-3}$~TeV$^{-1}$ (or, ideally, ALP discovery) relatively uniformly over $m_a\approx 1$~GeV--5~TeV. Going to lower couplings for axion-like particles with masses in the multi-GeV regime, will require a next generation of $\epem$ colliders such as FCC-ee~\cite{Abada:2019zxq}, and ILC~\cite{Steinberg:2021iay}.


\end{document}